\documentclass[12pt]{article}

\usepackage{scicite}
\usepackage{times}

\usepackage{color,gensymb,amsmath,amssymb,graphicx,morefloats}
\usepackage{bm} 
\usepackage{graphicx,morefloats}
\usepackage[]{multibib}
\usepackage{subfigure} 
\usepackage{color,soul}
\usepackage{caption}
\newcommand{\CPS}{Ce$_3$Pd$_{20}$Si$_6$}

\newcommand{\ve}[1]{\boldsymbol{#1}}

\newcites{SM}{Supplementary References}

\begin{document}

\thispagestyle{empty}

\hyphenation{va-ni-sh-ing de-lo-ca-li-za-tion}



\baselineskip24pt

\begin{center}

{\Large Quantum Fisher information in a strange metal}\\[0.4cm]

\normalsize{Federico Mazza$^{1,\ast}$, Sounak Biswas$^{2,\ast}$, Xinlin Yan$^1$, Andrey Prokofiev$^1$, Paul Steffens$^3$, Qimiao Si$^4$,
Fakher~F.~Assaad$^{2,5,\ddagger}$, and Silke Paschen$^{1,\ddagger}$}\\[0.4cm]

\small\it{$^1$Institute of Solid State Physics, TU Wien, Wiedner Hauptstr.\ 8-10, 1040 Vienna, Austria\\[0.1cm]

\small\it{$^2$Institut f\"ur Theoretische Physik und Astrophysik, Universit\"at W\"urzburg, 97074 W\"urzburg, Germany}\\[0.1cm]

\small\it{$^3$Institut Laue-Langevin, 71 Ave Martyrs, 38042 Grenoble 9, France}\\[0.1cm]

\small\it{$^4$Department of Physics and Astronomy, Center for Quantum Materials, 6100 Main Street, Rice University, Houston, Texas 77005, USA}}\\[0.1cm]

\small\it{$^5$W\"urzburg-Dresden Cluster of Excellence ct.qmat, 97074 W\"urzburg, Germany}

\vspace{0.4cm}

\today

\end{center}
\vspace{-0.2cm}

\noindent{\bf A strange metal is an exotic state of correlated quantum matter; intensive efforts are ongoing to decipher its nature. Here we explore whether the quantum Fisher information (QFI), a concept from quantum metrology, can provide new insight. We use inelastic neutron scattering and quantum Monte Carlo simulations to study a Kondo destruction quantum critical point, where strange metallicity is associated with fluctuations beyond a Landau order parameter. We find that the QFI probed away from magnetic Bragg peaks, where the effect of magnetic ordering is minimized, increases strongly and without a characteristic scale as the strange metal forms with decreasing temperature, evidencing its unusual entanglement properties. Our work opens a new direction for studies across strange metal platforms.}\\

\noindent$^{\ast}$These authors contributed equally to this work.


\noindent$^{\ddagger}$To whom correspondence should be addressed;\\
E-mails: paschen@ifp.tuwien.ac.at, fakher.assaad@uni-wuerzburg.de

\newpage

\noindent Strange metal behavior refers to a linear temperature dependence of the electrical resistivity at low temperatures instead of the square-in-temperature Fermi liquid form. First recognized in the cuprate high-temperature superconductors, strange metals are being identified in an increasing number of materials classes, from heavy fermion, pnictide, and organic compounds to frustrated-hopping and moir\'e flat band systems \cite{Che24.1}. Heavy fermion compounds have taken a prominent role in the search for other salient features of strange metallicity \cite{Pas21.1}, and a Fermi volume jump \cite{Pas04.1,Fri10.2}, dynamical scaling in the spin \cite{Sch00.1} and charge (or current) response \cite{Pro20.1}, and the suppression of shot noise \cite{Che23.1} have been put forward. They are all consistent with the static Kondo screening transitioning, in the zero-temperature limit, from being in place to being absent \cite{Si01.1,Col01.1,Sen03.1}, a scenario that is actively pursued with various theoretical techniques \cite{Dan20.1,Wan20.2,Cai20.2,Gle23.1x}. With accumulating evidence for these features in other strange metal platforms \cite{Che24.1} and theoretical efforts \cite{Ram21.1,Son22.1,Cho23.1,Che23.1x} to understand these systems in Kondo-based frameworks, the Kondo destruction (or breakdown) scenario may well be pertinent beyond the heavy fermion setting. However, very different scenarios are also considered \cite{Cho22.1,Phi22.1,Bas23.1x} and a unified understanding is still lacking.

Here we explore the potential of a new probe---the quantum Fisher information (QFI)---to make progress. As recently shown theoretically \cite{Hau16.1}, the QFI can be defined for condensed matter systems in thermal equilibrium via a Kubo response function

\vspace{-0.5cm}
\begin{equation}
f_{\rm Q}(T) = \frac{4}{\pi} \int_0^{\infty} \tanh{\left(\frac{\hbar\omega}{2 k_{\rm B}T}\right)}\chi''(\omega,T){\rm d}(\hbar\omega) \label{QFI} 
\end{equation}
involving the imaginary part of a dynamic susceptibility $\chi''(\omega,T)$, for instance the dynamic spin susceptibility that can be derived from inelastic neutron scattering (INS) experiments. In this formulation, called the QFI density \cite{Lau21.1andLau21.1err}, the susceptibility is an intensive quantity, i.e., it is counted per site or moment. The benefit of this new tool is that it extracts the entanglement content of the quantum correlations contained in $\chi''(\omega,T)$, and may thus provide complementary information to dynamical scaling analyses. A prediction of direct pertinence for the present work is that at ``strongly entangled'' quantum phase transitions, the QFI is expected to diverge in the $T=0$ limit, whereas no signature is expected at a thermal phase transition \cite{Hau16.1}. By contrast, in a spin-chain material, enhanced values of the QFI were found to be tied to the N\'eel order parameter and to decrease as the order is suppressed \cite{Lau21.1andLau21.1err}. Motivated by the finding that, at the Kondo destruction quantum critical point of a Kondo impurity model, the entanglement entropy becomes long-ranged \cite{Wag18.1} we set out to study a strange metal heavy fermion compound by INS experiments. We find that the QFI associated with the fluctuations that govern the strange metallicity, and go beyond order parameter fluctuations, increase strongly with decreasing temperature as the strange metal develops, evidencing a state with enhanced entanglement.

\begin{figure}[t!]
\vspace{-2cm}

\centering{\includegraphics*[width=0.85\textwidth]{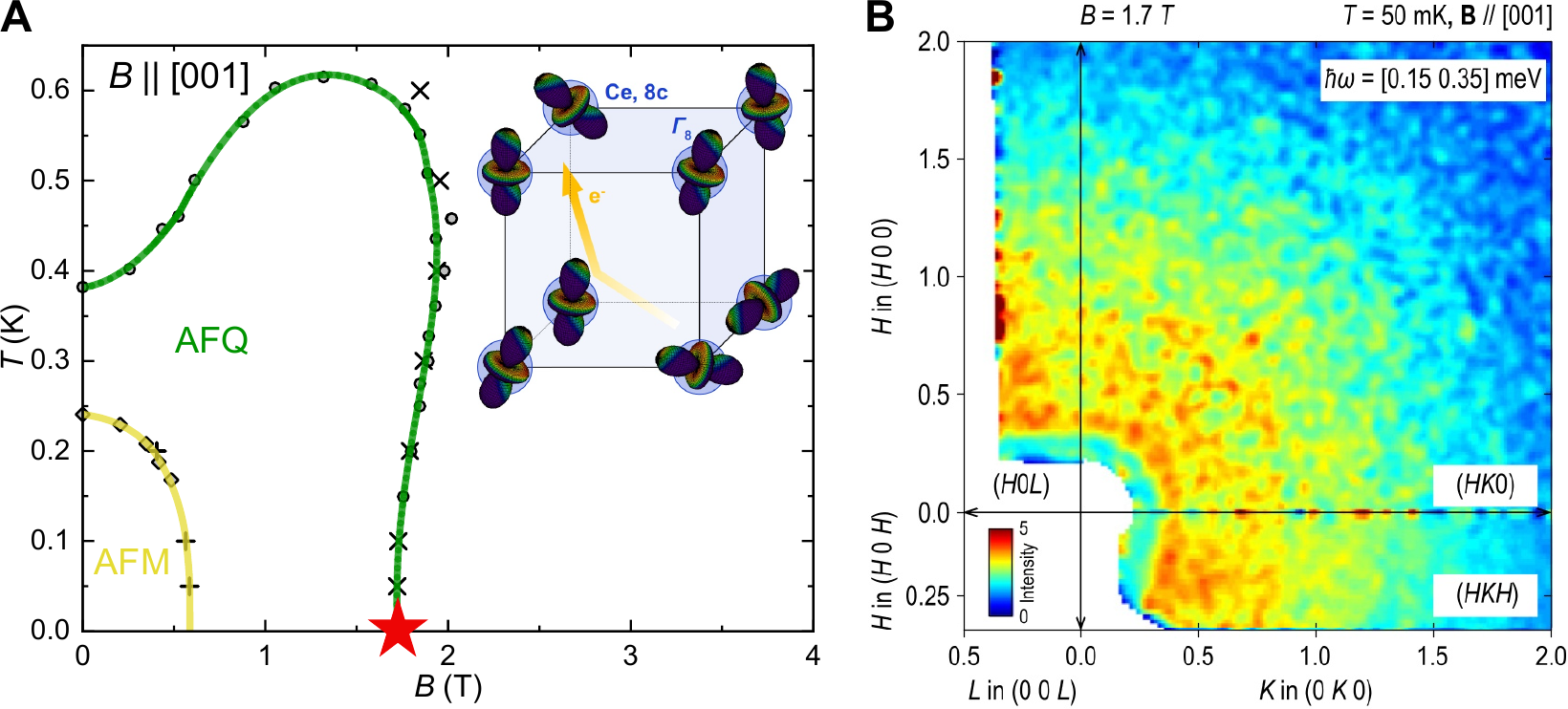}}
\vspace{0.3cm}

\centering{\includegraphics*[width=0.85\textwidth]{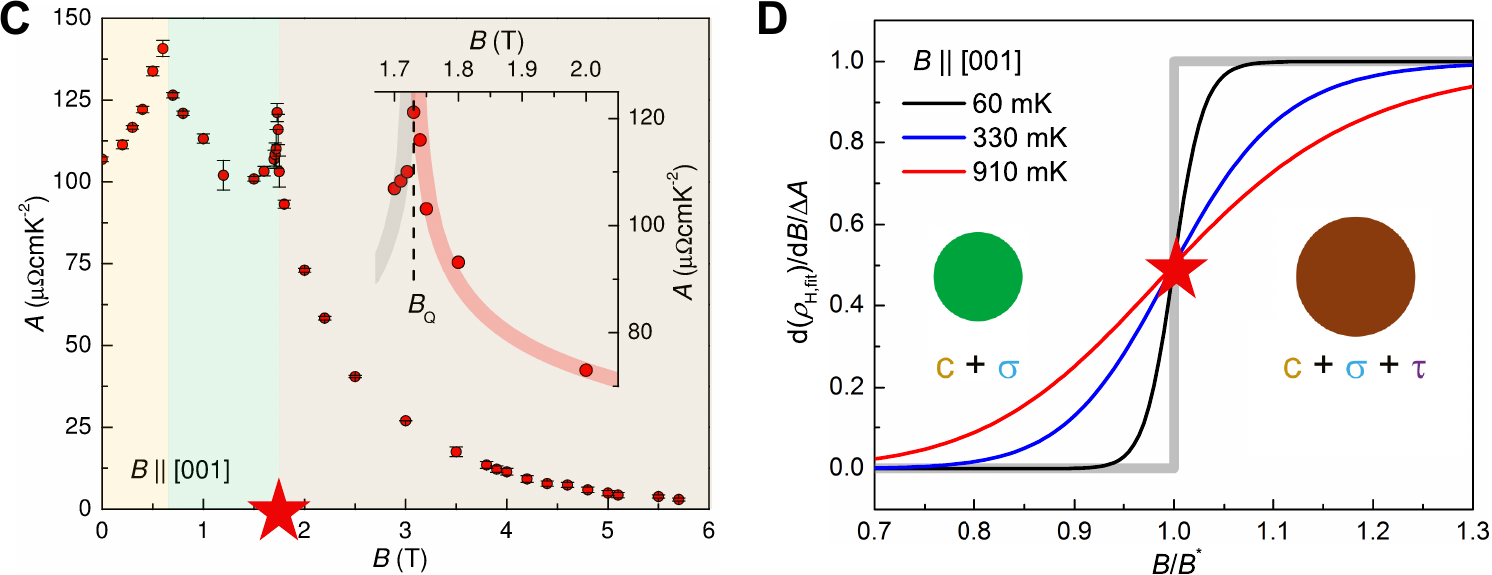}}
\vspace{0.3cm}

\caption{\label{fig1} {\bf The heavy fermion compound Ce$_3$Pd$_{20}$Si$_6$, with orbital moments undergoing Kondo destruction.} ({\bf A}) Cartoon of the crystal structure, showing only the magnetically active Ce atoms at the $8c$ positions with their $4f$ orbitals, which assume a $\Gamma_8$ quartet ground state. ({\bf B}) Constant-energy map at a magnetic field of 1.73\,T applied along the crystallographic $[0\,0\,1]$ direction and at 50\,mK, obtained by integrating time-of-flight data within the indicated energy range, and within $\pm 0.08$\,r.l.u.\ in the orthogonal momentum direction \cite{Por19.1}. The broad distribution of intensity in momentum space indicates the local nature of the fluctuations in real space. As shown previously, magnetic field applied along $[0\,0\,1]$ drives the material across two Kondo destruction quantum critical points: one at $B_{\rm M} \sim 0.7$\,T \cite{Cus12.1} where AFM order with the incommensurate ordering wavevector $(0\,0\,0.8)$ is suppressed \cite{Por16.1}, and one at $B_{\rm Q} \sim 1.73$\,T \cite{Mar19.1} where antiferroquadrupolar order with the ordering wavevector $(1\,1\,1)$ is suppressed \cite{Por16.1}. The latter one is studied here. ({\bf C}) The electrical resistivity follows the Fermi liquid form $\rho = \rho_0 + A T^2$ at the lowest temperatures, in shrinking temperature ranges upon approaching the critical fields, and with a strongly enhanced $A$ coefficient upon approaching those fields, consistent with divergences as shown for fields near $B_{\rm Q}$ in the inset. At both critical fields and in quantum critical fans emerging from them, the resistivity assumes the strange metal form $\rho = \rho_0' + A' T$ (see \cite{Mar19.1}). ({\bf D}) The differential Hall resistance jumps in the extrapolated zero-temperature limit \cite{Mar19.1}, both at $B_{\rm Q}$ as shown here and at $B_{\rm M}$ (see \cite{Cus12.1}). This is understood as resulting from the spin degree of freedom $\sigma$ of the $\Gamma_8$ quartet being incorporated into the Fermi surface at $B_{\rm M}$, and the orbital degree of freedom $\tau$ being incorporated at $B_{\rm Q}$ \cite{Mar19.1,Cus12.1}, via the Kondo destruction (or {\em con}struction) mechanism. Panel B is adapted from \cite{Por19.1}, panels C and D from \cite{Mar19.1}.}
\end{figure}

We chose the heavy fermion metal \CPS\ for this study because quantum criticality of Kondo destruction type, associated with strange metal behavior, has been identified in previous experiments \cite{Mar19.1}. We focus on the material's quantum critical point (QCP) near 1.73\,T (applied along $[0\,0\,1]$), where antiferroquadrupolar (AFQ) order is continuously suppressed (Fig.\,\ref{fig1}A). The absence of magnetic dipolar order near this QCP helps to reduce the effect of magnetic order parameter fluctuations, which are in general not expected to lead to strange metallicity \cite{Ros99.1}. As a second step towards this goal, we have selected the wave vector $(0\,\bar{1}\,0)$ for our INS study, which corresponds to neither a structural nor a magnetic Bragg peak \cite{Por16.1,Por19.1}. A characteristic of Kondo destruction quantum criticality is that the associated fluctuations spread out in momentum space, as seen at the QCP near 1.73\,T in \CPS\ \cite{Por19.1} (Fig.\,\ref{fig1}B). The fluctuations probed at $(0\,\bar{1}\,0)$ should thus be representative of the Kondo destruction fluctuations across the entire Brillouin zone (BZ).

As shown previously \cite{Mar19.1}, magnetic field applied along $[0\,0\,1]$ drives the material across a two-stage Kondo destruction transition \cite{Liu23.1,Sch23.1}. At large fields, both the spin and the orbital degree of freedom of the $4f^1$ $\Gamma_8$ quartet ground state of the magnetically active Ce atoms situated at the $8c$ site (inset of Fig.\,\ref{fig1}A) are Kondo screened. With decreasing magnetic field, at $B_{\rm Q} \sim 1.73$\,T, Kondo screening first breaks up for the quadrupole moments, leading to AFQ order with the ordering wavevector $(1\,1\,1)$ \cite{Por16.1}. With further decreasing field, at $B_{\rm M} \sim 0.7$\,T, Kondo screening breaks up for the spin degree of freedom \cite{Cus12.1}, leading to antiferromagnetic (AFM) order with the incommensurate ordering wavevector $(0\,0\,0.8)$ \cite{Por16.1}. Near both critical fields, the effective mass as probed by the $A$ coefficient of the Fermi liquid form $\Delta \rho = A T^2$ is strongly enhanced (Fig.\,\ref{fig1}C) before, at the two QCPs, the strange metal linear-in-temperature form prevails down to the lowest temperatures \cite{Mar19.1}. Both QCPs were also shown to feature the Hall effect characteristics of a Fermi surface jump \cite{Cus12.1,Mar19.1} (see Fig.\,\ref{fig1}D for the one at $B_{\rm Q}$).

\begin{figure}[ht!]
\vspace{-1cm}

\centering{\includegraphics*[width=0.85\textwidth]{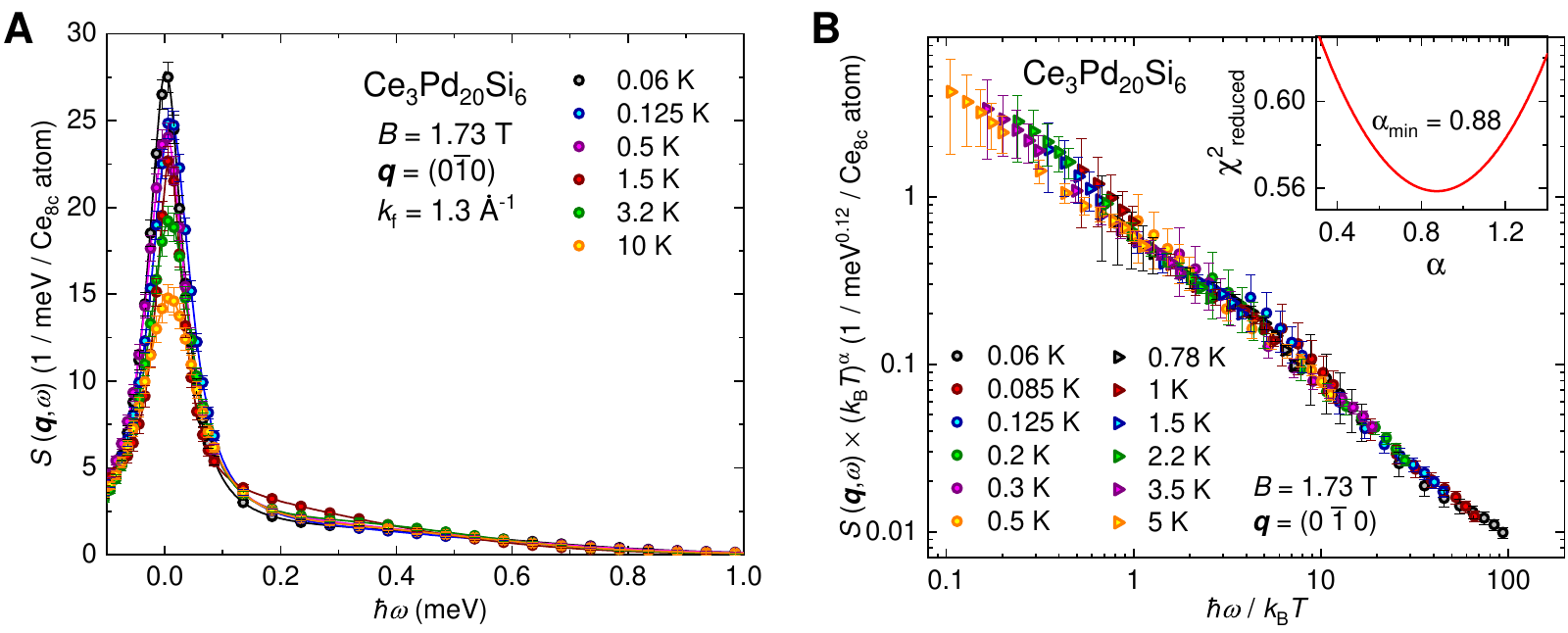}}
\vspace{0.3cm}

\caption{\label{fig2} {\bf Dynamic spin correlation function and dynamical scaling analysis of Ce$_3$Pd$_{20}$Si$_6$.} ({\bf A}) Selected isotherms of the dynamic spin correlation function $S(\bm{q},\omega,T)$ vs energy $\hbar\omega$, measured at $\bm{q} = (0\,\bar 1\,0)$ and in a magnetic field of 1.73\,T applied along $[0\,0\,1]$. ({\bf B}) $S(\bm{q},\omega,T)$ from (A), multiplied with $k_{\rm B}T^{\alpha}$ and plotted vs $\hbar\omega/k_{\rm B}T$. Data in the temperature range 0.06 - 5\,K and for energy transfers in the range 0.025 - 0.58\,meV show best overlap for the exponent $\alpha = 0.88$, as seen from the minimum in the quality factor $\chi^2$ of the minimization procedure ({\bf inset}). This scaling is compatible with Kondo destruction quantum criticality as described in \cite{Si01.1}.}
\end{figure}

We now turn to the INS data of \CPS. The experiment was performed at the cold-neutron triple-axis spectrometer ThALES (ILL, Grenoble), which has state-of-the-art energy resolution of $\sim 0.07$\,meV for the chosen final neutron wave vector $k_{\rm f}$, down to temperatures of 60\,mK. Extreme care was taken to remove all background contributions and to bring the data into absolute units (Supplementary Materials). The thus obtained dynamic spin correlation function $S({\bm q},\omega,T)$, measured at ${\bm q} = (0\,\bar 1\,0)$, is shown in Fig.\,\ref{fig2}A. At a Kondo destruction quantum critical point, the imaginary part of the dynamical spin susceptibility $\chi''({\bm q},\omega,T)$ is expected \cite{Si01.1} to exhibit the form

\vspace{-0.5cm}
\begin{equation}
\chi''({\bm q},\omega,T) = \frac{1}{A T^{\alpha} W ( \hbar\omega/k_{\rm B}T )} \; , \label{scaling} 
\end{equation}
at least at the ordering wave vector. Here we probe it at ${\bm q} = (0\,\bar 1\,0)$ which, as discussed above, is away from any magnetic (and structural) Bragg peak. $\chi''({\bm q},\omega,T)$ is related to $S({\bm q},\omega,T)$ via the fluctuation-dissipation theorem \cite{Men23.1,Sch21.1andSch21.1err} as

\vspace{-0.5cm}
\begin{equation}
\chi''({\bm q},\omega,T) = \pi (1-e^{-\hbar\omega/k_{\rm B}T})S({\bm q},\omega,T)\; . \label{fluct} 
\end{equation}
A minimization procedure of our $S({\bm q},\omega,T)$ data for energies below 0.58\,meV and temperatures below 5\,K yields the best data collapse for the exponent $\alpha = 0.88$ (Fig.\,\ref{fig2}B). The excellent quality of the scaling, together with the fractional exponent $\alpha$, provides strong evidence for the beyond-order-parameter nature of the quantum criticality. Given the broad intensity distribution of the inelastic scattering intensity across the BZ (Fig.\,\ref{fig1}B) we expect the same scaling to exist throughout the BZ, though a proof will require dedicated high-resolution (3-axis) INS experiments at other wave vectors.

Next, we determine the temperature-dependent QFI density

\begin{figure}[ht!]
\vspace{-1cm}

\centering{\includegraphics*[width=0.5\textwidth]{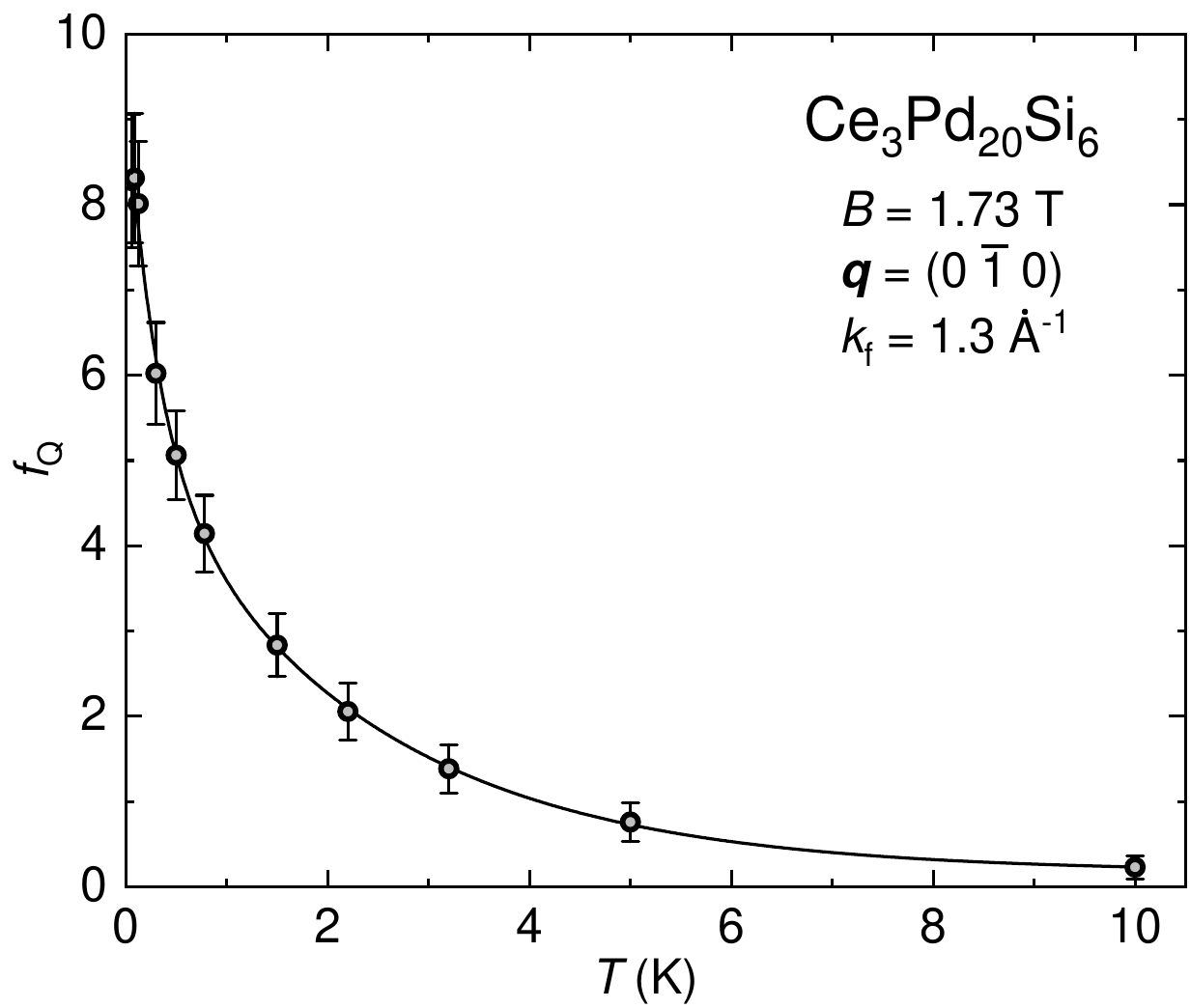}}
\vspace{0.3cm}

\caption{\label{fig3} {\bf Quantum Fisher information density of Ce$_3$Pd$_{20}$Si$_6$.} The data correspond to the ones presented in Fig.\,\ref{fig2}A, but all measured isotherms were analysed and the entire accessed energy range from 0 to 1.5\,meV was used for the integration.}
\end{figure}

\vspace{-0.5cm}
\begin{equation}
f_{\rm Q}(T) = 4 \int_0^{\infty} \tanh{\left(\frac{\hbar\omega}{2 k_{\rm B}T}\right)}(1-e^{-\hbar\omega/k_{\rm B}T})S(\omega,T){\rm d}(\hbar\omega) \label{QFId} 
\end{equation}
from the different $S({\bm q},\omega,T)$ isotherms at ${\bm q} = (0\,\bar 1\,0)$. $f_{\rm Q}$ shows a pronounced increase with decreasing temperature (Fig.\,\ref{fig3}), indicating that entanglement is building up as the Kondo destruction QCP is approached. The temperature dependence is smooth, without any sign of a characteristic energy scale or a trend of saturation. At the lowest accessed temperature of 60\,mK, $f_{\rm Q}$ reaches a value of 8.2.

To evaluate the entanglement depth associated with this value, one has to specify the type of interaction between the neutron and the sample. For scattering from localized spins, with a minimum and maximum expectation value $h_{\rm min}$ and $h_{\rm max}$ of the operator appearing in $S({\bm q},\omega,T)$, the system must be at least $(m + 1)$-partite entangled if $f_{\rm Q}$ satisfies the bound $f_{\rm Q} > m(h_{\rm max}-h_{\rm min})^2$, where $m$ is an integer  \cite{Hyl12.1,Hau16.1}. The normalized QFI \cite{Sch21.1andSch21.1err}

\vspace{-0.5cm}
\begin{equation}
{\rm nQFI} = \frac{f_{\rm Q}}{(h_{\rm max}-h_{\rm min})^2} \label{nQFI} 
\end{equation}
thus witnesses at least $(m+1)$-partite entanglement if ${\rm nQFI}>m$. Previous work has focused on the case of localized spin 1/2 systems \cite{Sch21.1andSch21.1err,Lau21.1andLau21.1err}, where $(h_{\rm max} -h_{\rm min})^2 = c g^2[(+1/2)-(-1/2)]^2=c g^2$. $c$ counts the spin directions that are probed by $S({\bm q},\omega,T)$ \cite{Xu13.2,Lau21.1andLau21.1err} and $g$ is the Land\'e factor.

In the case of \CPS, the quantum criticality we study arises from the suppression of an AFQ phase, which represents the order of electric quadrupoles arising in the spin-Kondo screened $\Gamma_8$ quartet \cite{Mar19.1}. A priori, neutrons do not couple to electric quadrupoles. However, a magnetic field can induce magnetic dipoles (along its direction) as secondary moments $\mu_{\sec}$. These are generally considered to be very small compared to a corresponding primary moment $\mu$ \cite{Mat12.2}. Any ratio $r=\mu_{\sec}/\mu<1$ will boost ${\rm nQFI}$ as ${\rm nQFI}/r^2$. For \CPS, with a Land\'e factor $g=1$ \cite{Maz22.1}, $c=1$ because the secondary moments are $B$-induced, and an estimated ratio $r=0.1$ (Supplementary Materials) we obtain ${\rm nQFI} = 8.2/(1 \times 1^2)/(0.1^2) = 820$. The very most conservative estimate (using $r=1$) is ${\rm nQFI} = 8.2$, but much larger values are likely. Additional experiments are needed to determine the actual value of $r$.

In what follows we describe auxiliary field quantum Monte Carlo simulations of a Kondo destruction transition and compare them to our experimental results. These simulations are unbiased and, most importantly, allow us to obtain wave-vector-resolved information. As a (sign-problem-free) model, we use a spin 1/2 Heisenberg chain on a two-dimensional Dirac semimetal akin to graphene \cite{Dan20.1}. The exchange interaction among the local moments of the spin chain competes with the Kondo coupling $J_{\rm K}$ of the local moments to the conduction electrons, which possess a pseudogap \cite{Wit90.1}. In the Kondo screened phase, a new particle described by the composite fermion operator $\hat{\Psi}^{\dagger}_{\ve{i},s} $ \cite{Dan21.1} emerges. It carries the quantum numbers of the electron and participates in the Luttinger volume such that this state can be identified as the heavy fermion phase with large Fermi surface. In the Kondo destruction phase at low $J_{\rm K}$, the composite fermion spectral function is purely incoherent (Supplementary Materials). The transition between these two regimes is driven by charge degrees of freedom and a sudden change in the Luttinger volume count at zero temperature.

\begin{figure}[ht!]
\vspace{-1cm}


\centering{\includegraphics*[width=0.85\textwidth]{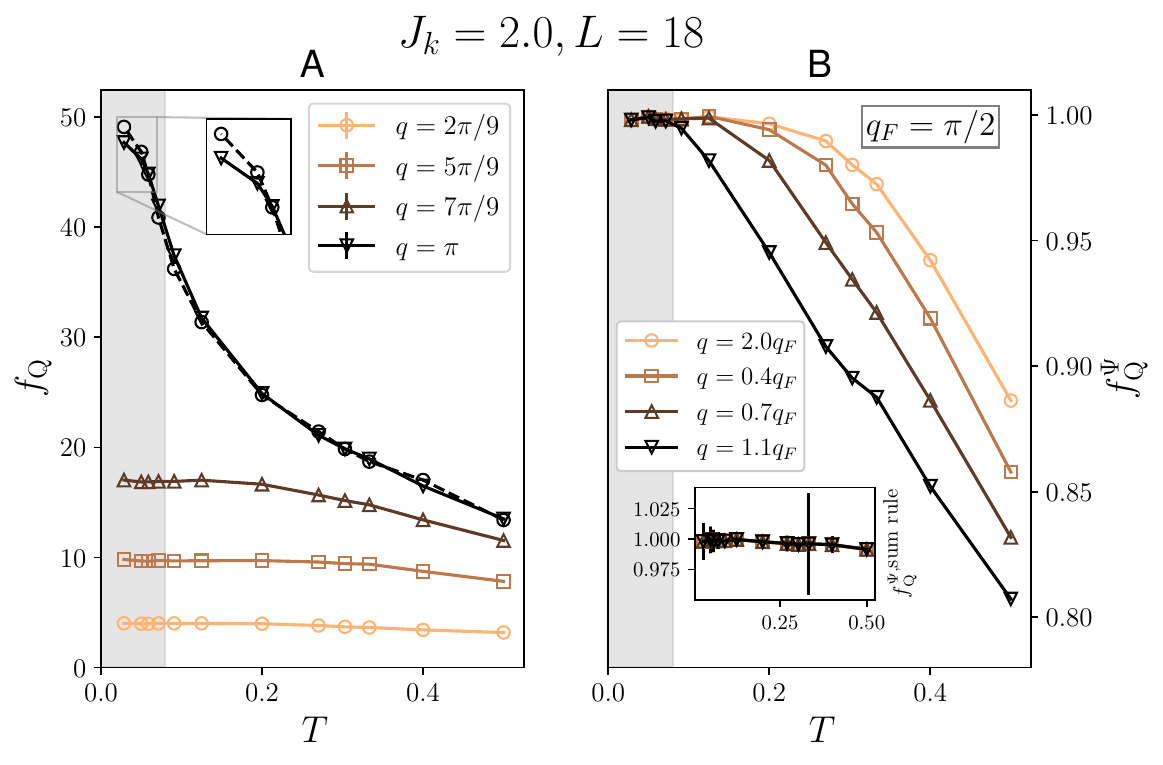}}
\vspace{0.3cm}

\caption{\label{fig4} {\bf Quantum Monte Carlo simulations of the QFI at the Kondo destruction transition.} ({\bf A}) QFI density $f_{\rm Q}$ for the spin degree of freedom at the ordering wave vector $q=\pi$ and at several wave vectors away from it. At the ordering wave vector, $f_{\rm Q}$ grows substantially in the low-temperature limit, but at other vectors, it converges to a finite value. The shaded region indicates where effects due to finite system size in. The dashed curve is for a larger system ($L=22$) and tends to saturate only at lower temperatures. As our $f_{\rm Q}$ takes into account all three components of spin-spin correlations, we have $(h_{\rm max}-h_{\rm min})^2=3g^2=12$ in Eq.~\ref{nQFI}, and hence ${\rm nQFI}=f_{\rm Q}/12$. ({\bf B}) QFI density $f^{\Psi}_{\mathrm Q}(T)$ for the composite fermion. Wave vectors closer to the Fermi surface $q_{\rm F}=\pi/2$ show the most pronounced temperature dependence reflecting the spectral weight corresponding to the heavy fermion quasiparticle. The shaded area has the same meaning as in (A). \textbf{Inset}: The wave-vector-independent and weakly temperature-dependent contribution to $f^{\Psi}_{\mathrm Q}(T)$ from the sum rule (Supplementary Materials).}
\end{figure}

The differences between the model and material are obvious, and what we aim for is to understand which aspects are generic and which are model-specific. We first describe the results for the QFI in the spin channel, obtained at the ordering wave vector $q = \pi$ of the chain. In the Kondo destruction phase, $f_{\rm Q}$ is that of an isolated spin 1/2 chain and diverges as $\ln(1/T)^{3/2}$ \cite{Men23.1}. In the Kondo screened phase, the spin degrees of freedom become itinerant, and $f_{\rm Q}$ saturates to a finite value as for a Fermi liquid state (Supplementary Materials). At the Kondo destruction QCP, by contrast, the QFI grows without characteristic temperature scale (Fig.\,\ref{fig4}A). We argue that this is the signature of Kondo destruction quantum criticality. $f_{\rm Q}$ reaches large values (which are limited by the finite system size of our simulations, see shaded region in Fig.\,\ref{fig4}A). Both the increase without a characteristic scale and the much-enhanced value of $f_{\rm Q}$ compared to the heavy fermion phase are thus common characteristics of our simulations and experiments. The large value of $f_{\rm Q}$ reflects the destruction of the composite fermion operator and the concomitant liberated critical spin degrees of freedom. For wave vectors away from $q = \pi$, a saturation of the QFI is seen even at the Kondo destruction QCP. This discrepancy to the experiment is due to the absence of ``local'' quantum criticality in the model. 

Before describing the results for the composite fermion operator, a few comments on the Kondo destruction side of the QCP are due. Here, the spin degrees of freedom can form various phases of matter, both with and without long-range order \cite{Pas21.1,Fuh21.1}, but this does not change the behavior at the QCP. For instance, changing the symmetry from SU(2) to SU(4) (antisymmetric self-adjoint representation) in our model calculations will frustrate the ordering, but should not alter the overall behavior of the QFI at criticality in both the spin and single particle channels. A recent extended dynamical mean-field theory (EDMFT) treatment of another model that features a Kondo destruction QCP shows a peak of the QFI at the QCP \cite{Fan24.1x}. The rise of $f_{\rm Q}(T)$ without a scale \cite{Fan24.1x} is similar to what we find here. We note that in the limit $T\rightarrow 0$, the $q$-resolved QFI converges to the static structure factor $4 S(q)$ with the sum rule $\frac{1}{N}\sum_q 4 S(q) = 4 \hbar^2 g^2 s(s+1)$ and $s=1/2$. This implies that if one wave vector dominates the low-temperature QFI, then the response at the others will be suppressed. Conversely, for a local quantum critical point with a fractional dynamical exponent \cite{Si01.1}, the QFI in the spin channel does not diverge \cite{Fan24.1x}.

We now show that the QFI of the composite fermion operator $\hat{O}_q = \sum_{s} \left(\hat{\Psi}^{\dagger}_{q,s} + \hat{\Psi}_{q,s}\right)$ captures the essence of the Kondo destruction transition. Being a fermion operator, its QFI density $f^{\Psi}_{\mathrm Q}$ is related to the single-particle spectral function $A(q,\omega)$ of the composite fermion via

\vspace{-0.5cm} 
\begin{equation}
\label{eq:FisherPsi}
 f^{\Psi}_{\mathrm Q}(T) = 2 \int_{-\infty}^{\infty} \tanh^2\left(\frac{ \hbar \omega}{2 k_{\rm B} T} \right) A(q,\omega) d (\hbar \omega) \; . 
\end{equation}
$f^{\Psi}_{\mathrm Q}(T) $ is bounded by the sum rule $\int A(q,\omega) d\omega$ (Supplementary Materials and Ref.~\cite{Dan21.1}) where $A(q,\omega)$ is taken at $T=0$. This sum rule has no $q$-dependence and only weak temperature dependence (inset of Fig.\,\ref{fig4}B). Importantly, the linear-in-temperature corrections to the sum rule are set by the integrated spectral weight in a window $k_{\rm B}T$ around the Fermi energy. Even though this quantity is not ideal for detecting multi-partite entanglement---it is bounded for all wave vectors by a $q$-independent sum rule---it provides valuable low-energy information in the single particle channel. The temperature dependence of $f^{\Psi}_{\mathrm Q}$ is strong around the Fermi wave vector $q_{\rm F}=\pi/2$, the low-temperature saturation being again a finite size effect (Fig.\,\ref{fig4}B). At the Kondo destruction QCP, we conjecture $f^{\Psi}_{\mathrm Q}(T)$ to have a unique temperature dependence down to $T=0$, again free of any scale. This shows that the destruction of the composite fermion operator (probed by $f^{\Psi}_{\mathrm Q}$) and the critical fluctuations of the thereby liberated spins (probed by $f_{\mathrm Q}$) go hand in hand, and together characterize Kondo destruction quantum criticality. It will be of great interest to study these quantities in other models and with other techniques, to test their universality.

The same is true for experiments on other materials. We note that it is crucial to access the lowest possible energies at the lowest possible temperatures and perform careful background measurements for meaningful results (Supplementary Materials). Nevertheless, even the best experiment is performed at non-zero temperature and on a spectrometer with finite resolution. Thus, the experimentally extracted QFI is unavoidably a lower bound for the actual situation. Though inelastic X-ray scattering \cite{Hal23.1} and ARPES \cite{Mal23.1x}-based methods to probe the QFI have been theoretically proposed, both the energy resolution and the lowest accessible temperatures of these methods are orders of magnitude away from what can be achieved with state-of-the-art INS experiments.

To the best of our knowledge, our work evidences the strongest entanglement detected to date in any many-body quantum system, including the recently investigated triangular antiferromagnet KYbSe$_2$, which is considered a proximate quantum spin liquid \cite{Sch24.1}. Whether such massive multi-partite entanglement is a general property of strange metals is an important question that should be addressed by future experiments across the strange metal platforms \cite{Che24.1}. On the other hand, our result may have potential for quantum applications. For instance, genuine multi-partite entanglement is necessary to reach the maximum sensitivity in certain metrological tasks \cite{Hyl12.1,Tot12.1}.




\newpage

\noindent{\bf ACKNOWLEDGMENTS}

\noindent We thank Lei Chen, Yuan Fang, Philipp Hauke, Karsten Held, Markus Heyl, Dmytro Inosov, Bernhard Keimer, Yong-Baek Kim, Pontus Laurell, Mounica Mahankali, Julia Mathe, Stanislav Nikitin, Subir Sachdev, Allen Scheie, Alan Tennant, Peter Thalmeier, Giuseppe Vitagliano, Yiming Wang, Peter Zoller for useful discussions.

\noindent{\bf Funding:}
Financial support for this work was provided by the Austrian Science Fund (SFB F 86, Q-M\&S; Cluster of Excellence quantA, 10.55776/COE1; I5868-N, FOR 5249, QUAST), the European Research Council (ERC Advanced Grant 227378, CorMeTop), and the European Union's Horizon 2020 research and innovation programme (grant agreement No 824109 -- EMP). S.B. and F.F.A.\ acknowledge financial support from the DFG under the grant AS 120/16-1 (Project number 493886309) that is part of the collaborative research project SFB Q-M\&S funded by the Austrian Science Fund (FWF) F 86 as well as the W\"urzburg-Dresden Cluster of Excellence on Complexity and Topology in Quantum Matter ct.qmat (EXC 2147, project-id 390858490). Work at Rice has primarily been supported by the NSF Grant No.\ DMR-2220603, the AFOSR Grant No.\ FA9550-21-1-0356, the Robert A. Welch Foundation Grant No.\ C-1411, and the Vannevar Bush Faculty Fellowship ONR-VB N00014-23-1-2870. We gratefully acknowledge the Gauss Centre for Supercomputing e.V.\ (www.gauss-centre.eu) for funding this project by providing computing time on the GCS Supercomputer SUPERMUC-NG at the Leibniz Supercomputing Centre (www.lrz.de, project number pn73xu), as well as the scientific support and HPC resources provided by the Erlangen National High Performance Computing Center (NHR@FAU) of the Friedrich-Alexander-Universit\"at Erlangen-N\"urnberg (FAU) under the NHR project b133ae. NHR funding is provided by federal and Bavarian state authorities. NHR@FAU hardware is partially funded by the German Research Foundation (DFG) -- 440719683. The numerical calculations were carried out with the Algorithms for Lattice Fermions (ALF) library \cite{Ass22.1}. 
 

\newpage



\clearpage

\newpage


\clearpage

\newpage


\clearpage

\newpage


\end{document}